\documentclass[10pt,letterpaper,twocolumn]{article} %% two column, final layout

\usepackage{ol2}
\usepackage[draft]{hyperref}
\usepackage{amsmath}

\begin{document}

\twocolumn[ %% activate for two-column option

\title{The boundary force exerted on spatial solitons \\in cylindrical strongly nonlocal media}

%% For REVTeX it is possible to automate superscript and e-mail callouts with the superscriptaddress option; see REVTeX4 documentation.

\author{Qian Shou, Yanbin Liang, Qun Jiang, Yajian Zheng, Sheng Lan, Wei Hu$^{*}$ and Qi Guo$^{*}$}
\address{Laboratory of Photonic Information Technology, South China Normal University, Guangzhou, 510631, China}
\address{$^*$Corresponding author: huwei@scnu.edu.cn, guoq@scnu.edu.cn}

\begin{abstract}We investigate the propagation of the spatial soliton in
cylindrical strongly nonlocal media by a novel method of image beam
of light. The effect of the boundary on the soliton acting as the
dynamic force for the soliton steering is equivalent to the force
between the soliton beam and the image beam. The trajectory of the
soliton is analytically studied which is in good agreement with the
experimental results.
\end{abstract}

\ocis{190.0190}

 ] %% activate for two-column option

\noindent Nonlocal spatial solitons are extensively
investigated[1-16] since the pioneering work of Snyder and
Mithchell[1]. A topic that has captured a rising interest is the
interaction between solitons and the boundaries[4-8], especially in
lead glass. The group of Morderchai Segev found the boundary caused
elliptic solitons and vortex solitons in lead glass and discussed
the influence of the boundary force on the soliton trajectory[5,6].
In liquid crystal A. Alberucci et al demonstrated the power-depended
soliton repulsion at the boundary[7,8]. The nonlocal nature of lead
glass lies in the thermal optical nonlinearity (Poisson type) which
is intrinsically infinite without boundaries[5]. Therefore the
behavior of the nonlocal solitons in lead glass can be greatly
influenced by the remote boundary and one can easily obtain solitons
output controlled by asymmetric boundary forces[6]. In addition, the
problem of light trajectory is generally settled by the method of
equivalent particle theory and light ray equation[6,17] since the
beam width is quite less than the media size. The particle behavior
of light suggests that the soliton can be taken as a point light on
the cross section of the propagation media.

In this paper by analogy with the method of images in
electrostatics, we introduce a method of image beam of light to deal
with the problem of the boundary effect on the soliton trajectory in
cylindrical lead glass. The boundary force exerted on the soliton
can be equivalent to the force of a remote image point light to the
soliton. The experimental data of soliton steering can be fitted by
analytical solution in good agreement.

\begin{figure}[ht]
\includegraphics[width=9cm]{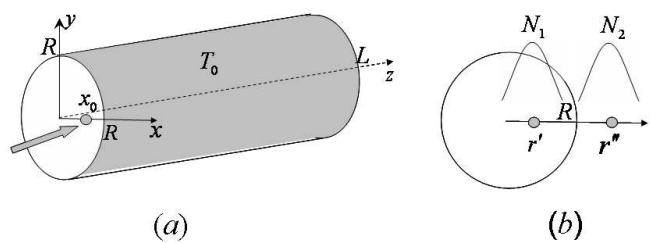}
\centering \caption{(a) Diagrammatic layout of a soliton launched at
an offset of $x_{0}$ in a circular cross section of the sample with
radius $R$. (b) Sketch map of the normalized refractive indexes
respectively induced by the source beam and the image beam. }
\end{figure}

The system we concern is the light-induced thermal self-focusing
nonlinearity in lead glass. The heat energy of the light is slightly
absorbed with an absorption coefficient $\alpha$ and diffused with a
thermal conductivity $\kappa$. The cylindrical boundary of the lead
glass is thermally contacted by a heat sink at a fixed temperature
$T_{0}$ described by Fig. 1(a). A temperature gradient is yielded
whose distribution is governed by Poisson equation[5]:
\begin {eqnarray}\label{1}
&&\nabla^{2}T(X,Y)= -\frac{\alpha}{\kappa} I(X,Y)
\nonumber\\&&T(X,Y)|_{X^{2}+Y^{2}=R^{2}}= T_{0},
\end {eqnarray}
where $T(X,Y)$ is the temperature distribution,
$I(X,Y)=|A(X,Y)|^{2}$ is the light intensity with $A(X,Y)$ being the
paraxial beam and $R$ is the radius of the cylinder cross section.
The heat transfer equation is inherently two-dimensional with
circular boundary as the soliton is invariant in the direction of
propagation. The laser-induced temperature change $\Delta T=T-T_{0}$
gives rise to a proportional increase in refractive index[5] $\Delta
n=\beta\Delta T$ with the thermo-optical coefficient $\beta$, and in
reverse the refractive index change has a strong impact on the
propagation characteristics of the light beam, including the
focusing and the steering. The propagation of an optical beam in
lead glass is moded by Equation (1) coupled with the equation for
the paraxial beam $A(X,Y)$:
\begin {equation}\label{2}
\nabla^{2}A+2ik\frac{\partial A}{\partial z}+2k^{2}\frac{\Delta
n}{n_{0}}A=0,
\end {equation}
where $k$ is the wave vector. We rewrite Eq. (1) and (2) in a
dimensionless form:
\begin {eqnarray}\label{3}
&&i\partial_{z} \varphi+\frac{1}{2}\nabla^{2}_{\bot}
\varphi+N\varphi=0 \nonumber\\&&
\nabla^{2}_{\bot}N=-|\varphi|^{2}\nonumber\\&&N(x,y)|_{x^{2}+y^{2}=1}=
0,
\end {eqnarray}
where the normalized coordinates and functions are setting as:
$x=X/R,y=Y/R,z=Z/(kR^{2}),\varphi=A/A_{0}$ with
$A_{0}^{2}=n_{0}\kappa/(\alpha\beta k^{2}R^{2})$ and
$N=k^{2}R^{2}\Delta n/n_{0}$. It is noticeable that $N$ and
$|\varphi|^{2}$ in Eq. 3(b) have their counterparts in
electrostatics, potential and charges respectively.

We solve the light-induced refractive index distribution in the view
of Green function method. In circular domain, the Green function of
Poisson equation can be deduced by the method of images[18] which is
the sum potential of the source charge and the image charge:
\begin {equation}\label{4}
G=\frac{1}{2\pi}({\rm ln}\frac{1}{|\mathbf{r}-\mathbf{r}'|}-{\rm
ln}\frac{r''}{|\mathbf{r}-\mathbf{r}''|})=G_{1}+G_{2},
\end {equation}
where $\mathbf{r}(x,y),\mathbf{r}'(x',y')$ and
$\mathbf{r}''(x'',y'')$ are respectively the normalized radius
vectors of the field point, the source point and the image point and
$r''=1/r'$ described in Fig. 1(b). The image charge simulates the
influence of all the inductive charges on the boundary, so the force
applied to the source charge by the boundary is equivalent to the
force between the image charge and the source charge. In our present
case, when the source light has a beam width small enough compared
with the boundary size, it can be taken as a point light beam.
Borrowing ideas from the method of images in the electrostatics we
introduce the method of image beam of light. By analogy with the the
expression of the charge interaction, we directly write the ``force"
between the source and image beams which provides the dynamic force
for the steering of the source beam:
\begin {equation}\label{5}
f=\frac{d^{2}x_{c}}{dz^{2}}\propto\frac{p}{1/x_{c}-x_{c}},
\end {equation}
where $(x_{c},0)$ is the source beam center, $(1/x_{c},0)$ is the
image beam center and $y$ has been set zero since the problem is
$y$-symmetric. $p=\int|\varphi(x'-x_{c},y')|^{2}dx'dy'$ is the
normalized light power which is equivalent to the charge density in
electrostatics. The $1/r$ law of interaction is ever predicted by C.
Rostschild et al[3].

The interaction of the two beams is mediated by the light-induced
index which is the solution of Eq. 3(c). When we make further
quantitative analysis, the solution is given by:
\begin {eqnarray}\label{6}
N&=&\frac{1}{2\pi}\int
(G_{1}+G_{2})|\varphi(x'-x_{c},y')|^{2}dx'dy'\nonumber\\&=&N_{1}+N_{2}.
\end {eqnarray}
The integration has been expressed in two terms of $N_{1}$ and
$N_{2}$. $N_{1}$ is symmetric about the beam center $(x_{c},0)$
since $G_{1}$ is shift invariant and the beam has a symmetric
profile(in the general case). It represents the source beam induced
refractive index in the free space who inversely serves as the
focusing lens for the source beam to form solitons. We can obtain
the critical power of the soliton $P_{c}=4\pi n_{0}\kappa/(\alpha
\beta k^{2}w_{0}^{2})$ with  $w_{0}$ being the beam width by
expanding $N_{1}$ to the second order[19]. The effect of $N_{2}$ on
the source beam is equivalent to the effect of the refractive index
induced by the image beam located at the image point $(1/x_{c},0)$
indicated in Fig. 1(b).

In the case of $w_{0}\ll R$, $G_{2}$ is almost unchanged in the
profile of the light beam, $N_{2}$ reads
\begin {eqnarray}\label{7}
N_{2}&=&-\frac{1}{2\pi}\int{\rm
ln}\frac{r''}{|r-r''|}|\varphi(x'-x_{c},y')|^{2}dx'dy'
\nonumber\\&=&-\frac{p}{2\pi}{\rm
ln}\frac{1/x_{c}}{\sqrt{(x-1/x_{c})^{2}+y^{2}}}.
\end {eqnarray}
The steering trajectory of a light ray is governed by the Eikonal
equation[9,17]:
\begin {equation}\label{8}
\frac{d^{2}x_{c}}{dz^{2}}=\frac{dN_{2}}{dx}|_{x=x_{c},y=0}=-\frac{p}{2\pi}\frac{1}{1/x_{c}-x_{c}}.
\end {equation}
Just as our qualitative anticipation in Eq. (5), the acceleration of
the source beam center has a form analogous to that of the
inter-force between the source and the image charges. We expand the
right hand side of Eq. (8) with respect to $x_{c}$ about $x_{c}=0$
under the condition of $x_{c}<<1$:
\begin {equation}\label{9}
\frac{d^{2}x_{c}}{dz^{2}}=-\frac{p}{2\pi}(x_{c}+x_{c}^{3}+x_{c}^{5}).
\end {equation}
We only provide the solution under the first-order approximation
since the solutions under the third and fifth-order approximations
containing Jacobi cn functions are too long to be showed:
\begin {equation}\label{10}
x_{c}^{(z)}=x_{c0}{\rm cos}(\sqrt{\frac{p}{\pi}}z),
\end {equation}
where $x_{c0}$ is the normalized input offset. The normalized output
coordinate $x_{c}^{(l)}$, where $l$ is the normalized length of the
lead glass, is proportional to $x_{c0}$ when the input light power
maintains constant.

Fig. 2(a) pictures the theoretical period of the beam steering
versus the input offset. The period decreases when the input offset
approaches to the boundary. This is understandable since strong
boundary effect accelerates the oscillation of the beam.
Nevertheless the oscillation period is too long compared to our
glass length even in the strong acceleration case. Fig. 2(b) is the
beam oscillation at the input offset of $0.8R$.

\begin{figure}[ht]
\includegraphics[width=8.5cm]{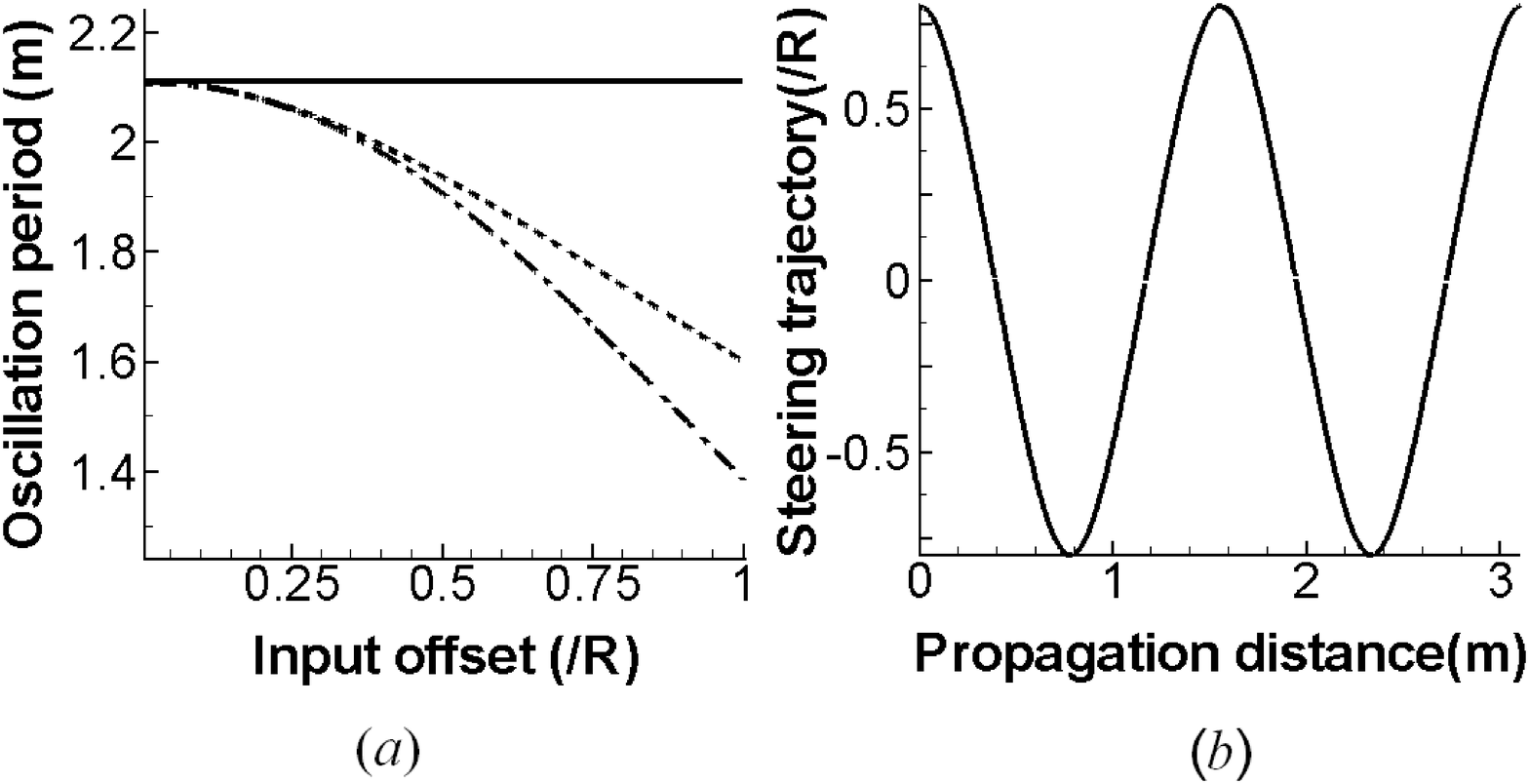}
\centering \caption{(a) The theoretical oscillation period verses
the input offset Solid line: first order approximation. Solid line:
first order approximation; dashed line: third order approximation;
dashdotted line: fifth order approximation. (b) The theoretical
period of the beam steering versus the input offset under the fifth
approximation.}
\end{figure}

Our experimental setput is similar to that for the work of B.
Alfassi et al[6]. The sample is a cylindrical heavily lead-doped
glass sample with the radius $R=1.5$ mm and length $L=5$ cm (Fig.
1(a)). The absorption coefficient $\alpha =0.07$ ${\rm cm}^{-1}$,
the thermo-optical coefficient $\beta =14 \times 10^{-6}$ ${\rm
K}^{-1}$, the refractive index $n_{0}=1.9$ and the thermal
conductivity $\kappa=0.7$ W/(mK). The 50 $\mu$m FWHM light input is
produced by a double frequency YAG laser(Verdi 5) with the
wavelength 532 nm. The soliton critical power is measured to be 500
$mW$ which is in good agreement with the analytical value[19] under
the experimental parameters. The data of the steering experiment are
obtained under the higher light power of 700 $mW$. Fig. 4 gives the
comparisons between the experimental data and the theoretical
fitting curves. The first approximation solution is a straight line,
it can be fitted to the experimental data only when the input offset
is small enough. The higher order approximation gives the better
agreement with the experimental result. It is noticeable that the
propagation distance in Fig. 2(b) is far longer than the actual
sample size. Yet we can not obtain large steering via lengthening
the glass or obtain the output at the desired period, because the
absorption will result in the changing of the beam width even the
collapse of the soliton.

\begin{figure}[ht]
\includegraphics[width=8cm]{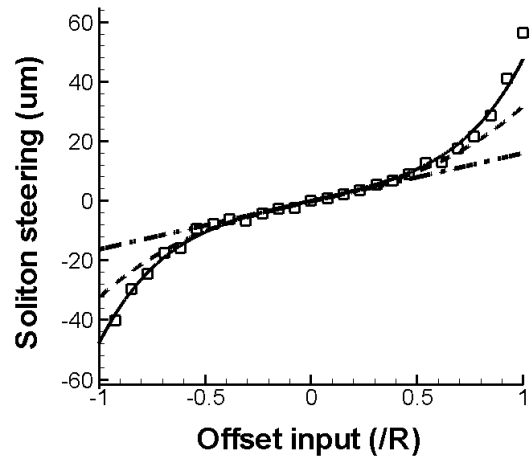}
\centering \caption{A net steering relative to the input offset
verses the input offset. Squares: the experimental result;
dashdotted line: the first approximation solution
$x_{c0}-x_{c}^{(1)}$; dashed line: the third approximation solution
$x_{c0}-x_{c}^{(3)}$; solid line: the fifth approximation solution
$x_{c0}-x_{c}^{(5)}$.}
\end{figure}

In conclusion, the soliton propagation in cylindrical lead glass is
studied. A novel method of image beam of light is produced by
analogy with the method of images in electrostatics. The boundary
effect on the soliton is treated as the inter-force between the
soliton beam and the image beam located at the image point. The
analytical solution of the soliton trajectory is in good agreement
with the experimental results. The method of image beam of light is
useful in the treatment of the boundary problems, including the
influence of the boundary force on the soliton trajectory and even
the interaction between solitons via the boundary effect.

This research was supported by the National Natural Science
 Foundation of China (Grant Nos. 10804033 and 10674050) and Program for Innovative Research Team of the Higher Education
 in Guangdong (Grant No. 06CXTD005).

%\pagebreak

\end{document}